\documentclass[12pt,reqno]{article}
\usepackage{amsmath}
\usepackage{amsthm}
\usepackage{amsfonts}
\usepackage{amssymb}
\usepackage{epic}
\usepackage{eepic}
\usepackage{epsfig}
\usepackage{graphics}
\input{xy}
\xyoption{all}

\newcommand{\bz}{\bar{z}}

\def\s{\sigma}

\def\bc{\bar\gamma}

\def\QR{\mathbb{R}} 
 
\def\QZ{\mathbb{Z}}
 
\def\a{\alpha}

\def\e{\epsilon}

\newcommand{\beq}{\begin{equation}}
\newcommand{\eeq}{\end{equation}}

\def\beas{\begin{eqnarray*}}
\def\eeas{\end{eqnarray*}}
\def\bea{\begin{eqnarray}}
\def\eea{\end{eqnarray}}
\def\a{\alpha}

\def\vp{\varphi}

\newcommand{\remlst}{\begin{list}
{(\arabic{num})}{\usecounter{num}\topsep0cm \itemsep0cm \parsep0cm}}

\newcommand{\eq}{\begin{eqnarray*}}
\newcommand{\qe}{\end{eqnarray*}}
\newcommand{\eqn}{\begin{eqnarray}}
\newcommand{\qen}{\end{eqnarray}}

\newcommand{\mat}{\begin{pmatrix}}
\newcommand{\tam}{\end{pmatrix}}

\def\g{\mathfrak{g}}

\def\H{{\cal H}}

\def\P{{\cal P}}

\newcommand{\ssum}{{\scriptstyle \sum}}

\def\SL2R{{SL(2,$\mathbb{R}$)}}
\def\SL2C{{SL(2,$\mathbb{C}$)}}
\def\SU2{{SU(2)}}

\def\sl2{{sl(2)}} 
\def\u11{{gl(1$|$1)}}
\def\U11{{GL(1$|$1)}}
\def\Mu11{{\mbox{\u11}}} 
\def\mU11{{\mbox{GL(1$|$1)}}}
\newcommand{\p}[1]{{\langle#1\rangle}}  
\def\J{{\cal J}}

\title{\bf The GL(1$|$1) WZW-Model: From \\[4mm] 
 Supergeometry to Logarithmic CFT\\[1cm]}  
\author{{\sc Volker Schomerus} and {\sc Hubert Saleur} \\[5mm] 
DESY Theory Group, Notkestrasse 85\\ 
D-22603 Hamburg, Germany. \\[5mm]
Service de Physique Th\'eorique, CEA Saclay,\\ 
F-91191 Gif-sur-Yvette, France;\\[5mm]
 Physics Department\\
University of Southern California\\
Los Angeles CA 90089-0484.\\
}
\date{October, 2005} 
\begin{document}
\begin{titlepage}      \maketitle       \thispagestyle{empty}

\vskip1cm
\begin{abstract} 
We present a complete solution of the WZW model on the supergroup 
GL(1$|$1). Our analysis begins with a careful study of its
minisuperspace limit (``harmonic analysis on the supergroup''). 
Its spectrum is shown to contain indecomposable representations. 
This is interpreted as a geometric signal for the appearance of 
logarithms in the correlators of the full field theory. We then 
discuss the representation theory of the gl(1$|$1) current 
algebra and propose an Ansatz for the state space of the WZW 
model. The latter is established through an explicit computation 
of the correlation function. We show in particular, that the 
4-point functions of the theory factorize on the proposed set 
of states and that the model possesses an interesting spectral 
flow symmetry. The note concludes with some remarks on 
generalizations to other supergroups. 
\end{abstract} 

\vspace*{-22cm}\noindent
{\tt {SPhT-T05/152}} \\
{\tt {DESY 05-186}} 
\bigskip\vfill 
\noindent
\phantom{wwwx}{\small e-mail: }{\small\tt  
volker.schomerus@desy.de, hubert.saleur@cea.fr } 
\end{titlepage} 

\baselineskip=19pt 

\section{Introduction} 

Throughout the last two decades, non-linear sigma models with 
super-manifold target spaces have emerged in a wide variety of 
systems and their study has become increasingly relevant for 
some of the challenging problems of modern physics, ranging 
from  e.g.\ the quantum Hall effect to the famous AdS/CFT 
correspondence in string theory. 
\smallskip 

In condensed matter, super-manifold target spaces arise mostly 
in the study of geometrical problems such as percolation and 
polymers \cite{Parisi}, or in non-interacting disordered systems 
\cite{Efetov:1983xg,Weidenmuller}, where ill defined $n\rightarrow 0$ 
``replica'' limits are handled instead by the introduction of 
fermionic degrees of freedom to, typically, cancel bosonic loops. The 
transition between plateaux in the integer quantum Hall effect is 
thus believed to be related to the sigma model U(1,1$|$2)$/$U(1$|$1)
$\times$ U(1$|$1) at $\theta=\pi$, a conformal field theory which has 
not yet been understood, despite decades of work (for a recent 
attempt, see \cite{Zirnbauer:1999ua}). Slightly more progress has 
been made for geometrical loop models, leading to partial solutions 
of sigma models on U(n+m$|$n)$/$U(1)$\times$ U(n+m-1$|$n) (super 
projective  spaces) and Osp(2n+m$|$2n)$/$Osp(2n+m-1$|$2n) 
(superspheres) \cite{Read:2001pz}.
\smallskip 
  
In string theory, super-manifold target spaces received brief 
attention about then years ago when they were argued to arise
as mirrors of rigid Calabi-Yau (CY) manifolds, i.e.\ of CY
spaces without complex moduli. According to the usual rules, 
the mirror image of such spaces has no K\"ahler moduli and 
hence it cannot be a usual CY manifold. In \cite{Sethi:1994ch} 
Sethi argued that the dual of a rigid CY is instead given by a CY 
super-manifold. The proposal was further investigated in a small
number of subsequent publications (see e.g.\ \cite{Schwarz:1995ak,
Aganagic:2004yh} and references therein), but it did not trigger 
much interest in sigma models with super-target spaces. Mirror
symmetry (or T-duality) involving non-commutative geometries, 
of which super-manifolds are the simplest examples, has also 
been discussed recently in \cite{Kapustin:2003sg,Mathai:2004qc,
Mathai:2004qq}. 
\smallskip 

Presumably more important, however, is the role that super-group 
and super-coset targets play for the description of strings 
in Anti-deSitter spaces. Using the Green-Schwarz formalism, a link 
was first established by Metsaev and Tseytlin \cite{Metsaev:1998it}. 
Shortly after, Berkovits, Vafa and Witten explained \cite{Berkovits:1999im} 
how string theory on $AdS_3 \times S^3$ could be quantized if it was 
possible to construct conformal quantum field theories with a PSL(2$|$2) 
target space. Such models were investigated in an interesting 
paper by Bershadsky \cite{Bershadsky:1999hk} in which some of 
the peculiar features of super-target spaces surfaced. For 
further string motivated research in this direction see e.g.\ 
\cite{deBoer:1999ie,Berkovits:1999zq,Bena:2003wd}, and for 
more condensed matter oriented work see \cite{Guruswamy:1999hi,
Bhaseen:1999nm,Saleur:2001cw,Essler:2005ag}). 
\medskip 

In most circumstances, the  models of interest are believed to 
be more complicated than WZW models on supergroups. In the case 
of the integer quantum Hall effect for instance, it has become 
clear over the years that the sigma model at $\theta=\pi$ 
flows to a theory which does not exhibit the full current algebra 
symmetry \cite{Read:2001pz} (presumably because of the appearance 
of logarithmic terms in the OPEs of the currents). Nevertheless, 
even the WZW models on supergroups are far from being understood. 
This is largely due to technical reasons (indecomposability of 
operator products and appearance of logarithms in correlation 
functions, continuous modular transforms of the irreducible 
characters \cite{Taormina}\ldots), combined with a lack of  
``physical intuition''.
\smallskip 

Our aim  in this note  is to initiate a systematic study of WZW 
models on supergroups by relying more heavily on  geometric
concepts. We shall, in particular, gain a better understanding of 
logarithmic features  by relating them to super-geometry. 
Logarithmic conformal field theories have been studied for 
a bit more than a decade now (see \cite{Rozansky:1992rx,
Gurarie:1993xq} for some early contributions). Even though
only a few examples have been constructed in full detail
\cite{Gaberdiel:1998ps}, 
their importance, in particular for disordered critical 
points, is widely appreciated (see e.g.\ \cite{Caux:1995nm,
Maassarani:1996jn,Read:2001pz,Gurarie:2004ce} and references 
therein). By definition, the operator product expansions in 
a logarithmic conformal field theory contain a logarithmic 
dependence on the separation between the fields. In the 
simplest cases, these may look e.g.\ as follows, 
$$ \Phi(x,\bar x)\, \Phi(0,0) \ \sim \ 
   \frac{1}{|x|^{2\Delta_\Phi-2h_C}}\, \left(\, \log|x|^2 
    C(0,0) + D(0,0)\, \right)\, +  \dots \ \ . 
$$ 
We conclude that the chiral generators $L_0$ and $\bar L_0$ 
of dilations in the world-sheet coordinate $x$ act according 
to    
$$ L_0 |D\rangle  \ = \ h_C |D\rangle + |C\rangle 
   \ \ \ , \ \ \ L_0 |C\rangle  \ = \ h_C|C\rangle
\ \ . $$ 
Here, $|C\rangle, |D\rangle$ denote the states that 
are associated with the fields $C, D$ and similar  
relations hold for $\bar L_0$. Consequently, $L_0$ and 
$\bar L_0$ cease to be diagonalizable. This feature is 
common to all logarithmic conformal field theories and 
it is rather easy to diagnose. Many more details and 
references may be found in recent review articles 
\cite{Gaberdiel:2001tr,Flohr:2001zs}. 
\smallskip 

Our strategy here is to approach the analysis of the 
WZW model through the harmonic analysis on the supergroup 
\U11 and to show that the minisuperspace analogues of 
$L_0$ and $\bar L_0$ , i.e.\ the quadratic Casimir 
elements in the left and right regular representations, 
are non-diagonalizable. This leaves the full field 
theoretic model no other chance but to be logarithmic. 
The harmonic analysis on the supergroup \U11 is the 
main subject of the next section. In section 3 we 
will suggest an expression for the state space of the 
full field theory. Our proposal is motivated in parts 
by the experience with the minisuperspace theory 
combined with some results from the representation 
theory of the \u11 current algebra. It is established 
later through a full construction of the theory, 
including all its correlators. Our solution is based 
on free field computations involving a $c=2$ linear 
dilaton in the bosonic sector and an (anti-)chiral 
bc-system with central charge $c = -2$ for the 
fermionic part. All 3-point functions of the model are 
constructed and studied in section 5. There we shall 
also show that the theory possesses an interesting 
spectral flow symmetry. In section 6, finally, we 
determine 4-point functions of our model an show that 
they factorize on the proposed set of states. We 
conclude with a few remarks on generalizations to other 
supergroups and with an outlook on further open problems.

\setcounter{equation}{0} 
\section{The minisuperspace analysis}

The following section is devoted to the ``particle limit'' of the 
\U11 WZW model .  In more physical 
terms, one can imagine putting the WZW model on a cylinder with 
periodic space and infinite (imaginary) time, and restricting to zero 
modes, ie to field configurations that are independent of the space 
variable. Their dynamics is the one of a particle with phase space the 
target space of the WZW model. Thus,  in more mathematical terms we shall be
concerned with the harmonic analysis on \U11.  Such harmonic analysis 
has been quite successful in the study of  WZW models on  non-compact 
bosonic target spaces such as the SL(2,C)/SU(2) model (see e.g. 
\cite{Teschner}). 

We  will 
require a bit of background from the 
representation theory of the Lie superalgebra gl(1$|$1). In 
particular we shall introduce its typical representations 
(long multiplets) and show how they generate certain 
indecomposable composites of atypical representations
(short multiplets) through tensor products. We then 
construct the space of functions on the supergroup along 
with the left and right regular action of gl(1$|$1). The  
regular representation is explicitly decomposed into its 
building blocks and it is shown that indecomposable (but 
not irreducible) representations emerge in the spectrum. 
Part of the results we discuss here were first derived 
in \cite{Huffmann:1994ah}.  

\subsection{The Lie superalgebra \u11 and its representations}

The Lie superalgebra $\g=$ \u11 is generated by two bosonic elements 
$E,N$ and two fermionic generators $\Psi^\pm$ such that $E$ is central 
and the other generators obey
$$  [N,\Psi^\pm] \ = \ \pm \Psi^\pm \ \ \ \mbox{and} \ \ \ \ 
    \{ \Psi^-,\Psi^+\} \ = \ E \ \ . 
$$ 
Let us also fix the following Casimir element $C$ for \u11 
$$ C \ = \ (2N-1) E + 2 \Psi^-\Psi^+ \ \ . $$
The choice of $C$ is not unique since we could add any 
function of the central element $E$. Our prescription is 
motivated by the form of the Virasoro element in the field
theory (see \cite{Rozansky:1992rx} and below).  
\smallskip 

There are five different classes of representations that shall 
play some role in the following. To begin with, we list the 
irreducible representations which fall into the different 
series. There is one series of 2-dimensional representations 
$\p{e,n}$ which is labeled by pairs $e,n$ with $e\neq 0$ and 
$n \in \QR$. In these representations, the generators take the 
form $E= e {\bf 1}_2$ and 
$$  N \ = \ \left(\begin{matrix} n-1 & 0 \\ 0 & n \end{matrix}\right) 
\ \ , \ \ 
 \Psi^+ \ = \ \left(\begin{matrix} 0 & 0 \\ e & 0 \end{matrix}\right) 
\ \ , \ \ 
 \Psi^- \ = \ \left(\begin{matrix} 0 & 1 \\ 0 & 0 \end{matrix}\right) 
\ \ . 
$$   
These representations are the typical representations (long 
multiplets) of $\g=$\u11. In addition, there is one series of 
atypical representations $\p{n}$ (short multiplets). These are 
1-dimensional and parametrized by the value $n \in \QR$ of $N$. 
All other generators vanish.  
\smallskip 

For the typical representations we assumed that the parameter 
$e$ does not vanish. But it is still interesting to explore 
what happens when we set $e=0$. The above matrices certainly 
continue to provide a representation of \u11 only that this 
is no longer irreducible. In fact, we observe that the basis
vector $|0\rangle = (1,0)^T$ generates a 1-dimensional 
invariant subspace of the corresponding 2-dimensional 
representation space. But one should not conclude that 
there exits an invariant complement. In fact, it is impossible 
to decouple the vector $|1\rangle = (0,1)^T$ from the representation 
since $\Psi^- |1\rangle = |0\rangle$, independently of the choice 
of the parameter $e$. The representation $\p{0,n}$ is therefore 
indecomposable but it is not irreducible. We can think of 
$\p{0,n}$ as being built up from two atypical constituents, 
namely from the representations $\p{n}$ and $\p{n-1}$. To 
visualize the internal structure of $\p{0,n}$, we may employ 
the following diagram, 
$$ \p{0,n}: \ \ \ \p{n-1} \ \longleftarrow\ \p{n} \ \ . $$
Later we shall see much more complicated composites of atypical 
representations. It is therefore useful to become familiar with 
diagrammatic presentations of indecomposables.%
\smallskip 

In the representations $\p{e,n}$, the fermionic generators appear 
on a somewhat different footing since $\Psi^+$ depends on the 
parameter $e$ while $\Psi^-$ does not. There exists another family 
of 2-dimensional representations $\p{\overline{e,n}}$, however, in 
which the roles of $\Psi^-$ and $\Psi^+$ are interchanged,    
$$  N \ = \ \left(\begin{matrix} n & 0 \\ 0 & n-1 \end{matrix}\right) 
\ \ , \ \ 
 \Psi^+ \ = \ \left(\begin{matrix} 0 & 1 \\ 0 & 0 \end{matrix}\right) 
\ \ , \ \ 
 \Psi^- \ = \ \left(\begin{matrix} 0 & 0 \\ e & 0 \end{matrix}\right) 
\ \ . 
$$  
As long as $e\neq 0$ the representations $\p{e,n}$ and $\p{\overline 
{e,n}}$ are equivalent. In fact, the isomorphism between the two 
representations may be implemented by conjugation with the matrices
$W_e = e\sigma^+ + \sigma^-$ where $\sigma^\pm$ are the usual Pauli
matrices. This isomorphism does not survive the limit $e\rightarrow
0$ and hence the representations $\p{0,n}$ and $\p{\overline{0,n}}$ 
are inequivalent. $\p{\overline{0,n}}$ is also an indecomposable 
representation that is built up from the same atypical constituents 
as $\p{0,n}$, but this time the non-vanishing generator $\Psi^+$ 
maps us from $\p{n}$ to $\p{n-1}$, i.e.\ 
$$ \p{\overline{0,n}}: \ \ \ \p{n-1} \ \longrightarrow\ \p{n} \ \ . $$   
Below, the representations $\p{0,n}$ and $\p{\overline{0,n}}$ will 
eventually enter as limits of typical representations.  
\smallskip

Having seen all the irreducible representations $\p{e,n}$ and
$\p{n}$ of \u11 along with their limits as $e$ goes to zero, 
our next task is to compute tensor products of typical 
representations $\p{e_1,n_2}$ and $\p{e_2,n_2}$. As long
as $e_1+e_2\neq 0$, the tensor product is easily seen to 
decompose into a sum of two typicals, 
$$ \p{e_1,n_2} \otimes \p{e_2,n_2} \ = \ \p{e_1+e_2,n_1+n_2-1} 
  \oplus \p{e_1+e_2,n_1+n_2}\ \ . $$
But when $e_1+e_2=0$ we obtain a 4-dimensional representation 
that cannot be decomposed into a direct sum of smaller 
subrepresentations. The representation matrices of these
4-dimensional indecomposables ${\cal P}_{n}$ read as 
follows
$$ N \ = \ \left(\begin{matrix} n-1 & 0 & 0 & 0 \\ 0 & n & 0 & 0 
 \\ 0 & 0 & n & 0 \\ 0 & 0 & 0 & n+1 \end{matrix}\right)  
\ \ , \ \ 
 \Psi^+ \ = \ \left(\begin{matrix} 0 & 0 & 0 & 0  \\ 
             -1 & 0 & 0 & 0 \\ 1 & 0 & 0 & 0 \\
                  0 & 1 & 1 & 0 \end{matrix} \right)
\ \ , \ \ 
 \Psi^- \ = \ \left(\begin{matrix} 0 & 1 & 1 & 0  \\ 
 0 & 0  & 0 & 1 \\ 0 & 0 & 0 & -1 \\ 0 & 0 & 0 & 0 
\end{matrix}\right) \ \ . 
$$   
As we have seen before, it is useful to picture the structure 
of indecomposables. The form of $N$ tells us that ${\cal P}_{n}$ 
is composed from the atypical irreducibles $\p{n-1},2\p{n}$, 
$\p{n+1}$. The action of $\Psi^\pm$ relates these four 
representations as follows
\begin{equation} 
{\cal P}_n: \ \xymatrix{ &  \p{n+1} \ar[dr] &  \\ 
\p{n} \ar[ur] \ar[dr] & & \p{n}\ . \\
& \p{n-1} \ar[ur] & } 
\end{equation}   
There are a few remarks we would like to make at this point. 
The first one concerns the form of the Casimir element $C$ in 
the representations ${\cal P}_n$. It is straightforward to see
that $C$ maps the subspace $\p{n}$ on the left onto the $\p{n}$ 
on the right of the above diagram and that it is zero otherwise.
This means that $C$ cannot be diagonalized in ${\cal P}_n$. We 
shall return to this observation later on. 
 
It is also obvious from the diagrammatic representation that 
${\cal P}_n$ contains the indecomposables $\p{0,n}$ and 
$\p{\overline{0,n+1}}$ as subrepresentations. In this
sense, the latter are extendable into a larger indecomposable. 
For the representation ${\cal P}_n$ the situation is quite 
different: it may be shown (and is intuitively clear) that 
${\cal P}_n$ is maximal in the sense that it can never appear 
as a subrepresentation of a larger indecomposable. In the 
mathematics literature, such representations are known 
as {\em projective}. Since the projective representation
${\cal P}_n$ contains the irreducible $\p{n}$ as a true
subrepresentation, one also calls ${\cal P}_n$ the 
{\em projective cover} of $\p{n}$. 
\smallskip 

The typical representations $\p{e,n}, e\neq 0,$ along with 
the indecomposables ${\cal P}_n$ exhaust the set of finite 
dimensional projectives of \u11. What will be particularly 
important for us is the fact that projective representations
are known to close under tensor products. In particular, 
tensor products of the representations ${\cal P}_n$ do 
not generate any new types of representations. This is not 
to say that there are not any others. In fact, there is 
a large family of indecomposables (``zigzag modules'')  with 
arbitrarily large dimension (see e.g.\ \cite{Gotz:2005jz} for 
a complete list and a computation of their tensor products). 
Our following analysis will shortly confirm the standard 
mathematical result that only projectives emerge from the 
harmonic analysis on the supergroup and hence these are 
the only ones that will play a major role below.

\subsection{Harmonic analysis on the supergroup GL(1$|$1)}  

Our aim now is to study the space of functions on the corresponding
supergroup and to analyse the various actions of the Lie superalgebra
\u11. Before we get into the details, however, let us briefly recall 
the situation in the case of compact groups which  is covered by 
Peter-Weyl theory. The latter describes how the space $L_2(G)$ of 
square integrable functions on a compact group decomposes under the 
right regular action. It asserts that the infinite dimensional 
representation space  $L_2(G)$ decomposes into a direct sum 
of irreducibles $H_J$ of $G$ and that each irreducible appears
with a multiplicity that is given by the dimension $d_J$ of 
$H_J$, i.e.\ 
$$ L_2(\mbox{G}) \ \cong \  \sum_J \ H_J \otimes H^R_J\ \ . $$  
Here, the first factor $H_J$ in each summand is the multiplicity 
space. The generators of the right regular representation act 
exclusively in the second tensor factor which is why we marked 
it with the superscript $R$. 
\smallskip 

We can actually be even more precise and construct each summand 
in the above decomposition rather explicitly. To this end we note 
that the Hilbert space $L_2(G)$ possesses a basis which is formed 
by matrix elements of irreducible representations of $G$. Any 
irreducible representation $H_J$ contributes $d_J^2$ matrix 
elements to the basis. These span the subspaces $H_J \otimes 
H^R_J$ in the above decomposition of $L_2(G)$. 
\smallskip 

Obviously, there exists a second action of $G$ on $L_2(G)$ by left 
multiplication. It promotes the multiplicity spaces $H_J$ into 
representation spaces of $G$, i.e.\     
$$ L_2(G) \ \cong \  \sum_J \ H^L_J \otimes H^R_J\ \ . $$  
The structure of this decomposition under the combined left 
and right action is somewhat reminiscent of the famous 
holomorphic factorization in WZW models. 
\smallskip

In the following discussion of functions on the supergroup, we 
would like to remain very explicit. Therefore, we introduce the 
so-called Gauss coordinates $x,y,\eta_\pm$ in which elements  
of the supergroup read 
$$ U \ = \ U(x,y,\eta_\pm) \ = \  
    e^{i\eta_+ \Psi^+} \, e^{ixE + iyN} \, e^{i \eta_-\Psi^-}
     \ \ .  
$$ 
It is not hard to work out the form of the invariant measure in these 
coordinates. The result is 
$$ d\mu \ = \ e^{-i y } dx\, dy\, d\eta_- d \eta_+ \ \ . $$ 
Similarly, one can determine the form of the left and right invariant 
vector fields. Again we only quote the results of a straightforward  
computation. For the left invariant vector fields one finds
\begin{equation}\label{lreg}  
   L_E \ = \ i \partial_x \ \ , \ \ 
   L_N \ = \ i \partial_y - \eta_+ \partial_+ \ \ , \ \ 
   L_+ \ = \ - i\partial_+ \ \ , \ \ 
   L_- \ = \  i e^{iy} \partial_- - \eta_+ \partial_x \ \ . 
\end{equation} 
Here, the symbols $\partial_\pm$ stand for derivatives with respect 
to $\eta_\pm$. Right invariant vector fields possess the form 
\begin{equation}\label{rreg}  
   R_E \ = \ - i \partial_x \ \ , \ \ 
   R_N \ = - i \partial_y + \eta_- \partial_- \ \ , \ \ 
   R_- \ = \ - i\partial_- \ \ , \ \ 
   R_+ \ = \ i e^{iy} \partial_+ +  \eta_- \partial_x\ \ .
\end{equation} 
The reader is invited to check that these two sets of generators 
satisfy the relations of the Lie superalgebra \u11 and that they 
(anti-)commute among each other. 
\medskip 

After this preparation, we would like to analyze the space of 
square integrable functions on the supergroup. By definition, 
these are objects $f$ of the form  
$$ f(x,y,\eta_\pm) \ = \ f_0(x,y) + f_+(x,y)\eta_+ + 
       f_-(x,y) \eta_- + f_2(x,y) \eta_-\eta_+ 
$$ 
with any set of square integrable functions $f_\nu$ on $\QR^2$. 
This space is spanned by the following basis\footnote{The 
elements of this basis are $\delta$-function normalizable 
since we are dealing with a non-compact group.}  
$$ e_0(k,l)  \ = \ e^{ikx+ily} \ \ , \ \ 
   e_\pm(k,l)  \ = \ e_0(k,l) \eta_\pm \ \ , \ \  
   e_2(k,l)  \ = \ e_0(k,l) \eta_- \eta_+ \ . 
$$ 
The space of square integrable functions carries two (anti-)commuting 
actions of the Lie superalgebra \u11 which are generated by the left- 
and right invariant vector fields. Our aim is to understand in detail 
the structure of these representations. 
\bigskip 

\noindent
{\bf Proposition 1:} (Right regular action) {\it  With respect to the 
right regular action, the space of square integrable functions 
on the supergroup decomposes according to 
$$ 
L_2(\mbox{\rm GL(1$|$1)}) \ = \ 
  \int_{e \neq 0} de dn \ \left( H'_\p{e,n} \oplus H^R_\p{e,n} 
   \right) \ \oplus \ 
   \int dn \ \P_n \ \ . 
$$ 
Here $H^R_\p{e,n}$ denotes the graded representation space of the 
typical representation $\p{e,n}$ and $H'_\p{e,n}$ is the same vector 
space with shifted $\QZ_2$ grading.} 
\bigskip 

Let us make a few remarks about this result before we explain its
derivation. The two integrals in our decomposition formula correspond
to an integration over the space of typical and atypical representations, 
respectively. As in the case of ordinary groups, typical representations
appear with a multiplicity given by their dimension, i.e.\ by $d_{\p{e,n}}  
= 2$ in our special case. For atypical representations, the story is more 
complicated. In general, they do not appear themselves but are replaced
by their projective covers. Their multiplicity, on the other hand, is
obtained from the dimension of the atypical representation, i.e.\ by 
$d_\p{n}= 1$ in our special case.\footnote{For some Lie superalgebras, 
there can be representations for which the multiplicity is only half 
of this value.} Our decomposition formula is thus in full 
agreement with the general result in \cite{Huffmann:1994ah}. Note 
that the structure of the sector which comes with the atypical 
representations does not possess the usual form that is encoded 
in the Peter-Weyl theorem (see above).

We also mention is passing that the Casimir element $C$ is 
non-diagonalizable in the right regular representation since 
the latter contains the projective covers and, according to 
our earlier discussion, $C$ cannot be diagonalized in 
${\cal P}_n$. In the full field theory, the Casimir 
elements lifts to the Virasoro zero mode $L_0$ so 
that our simple corollary on the structure of $C$ in 
the right regular representation will eventually have 
direct and far reaching implications for the WZW model. 
\smallskip 

Our result on the decomposition of the right regular representation 
is rather easy to obtain and we can even find explicit formulas for
the basis vectors of all the summands. In order to do so, we shall 
have a brief look at the space of functions that appear as matrix 
elements of the supergroup in the typical representations $\p{e,n}$, 
\begin{equation}\label{vpen}  
 \vp_{\p{e,n}} \ = \
   \left(\begin{matrix} e^{iex+i(n-1)y} 
    & i\eta_- e^{iex+i(n-1)y} \\ ie\eta_+ e^{iex+i(n-1)y} & 
    e \eta_- \eta_+ e^{iex+i(n-1)y} + e^{iex+iny}  
    \end{matrix}\right) \ \ .  
\end{equation} 
The functions in the first row form a basis of the summand 
$H_{\p{e,n}}$, whereas the functions in the second row span
the space $H'_{\p{e,n}}$, 
\begin{eqnarray*} 
H_\p{e,n} & = & \mbox{\rm span} \left(e_0(e,n-1),e_-(e,n-1)\right) 
\ \ \ , \\[2mm]      
H'_\p{e,n} & = & \mbox{\rm span} \left( e_+(e,n-1),  e e_2(e,n-1) 
            + e_0(e,n)\right)\ \ .  
\end{eqnarray*} 
It is obvious that the matrix elements of the typical representations
provide a basis for eigenfunctions of $R_E$ with eigenvalue $e \neq 0$. 
\smallskip

What we are missing is an analysis of the space of functions with 
$e=0$. The space of these functions is spanned by $e_\nu(0,l)$ and
it is easily seen to decompose into a sum of 4-dimensional 
indecomposables ${\cal P}_n$, 
\begin{equation} \label{Pj}  
 \P_n \ = \ \mbox{\rm span} \left( e_0(0,n), e_+(0,n-1), 
   e_-(0,n), e_2(0,n-1)\right) \ \ .     
\end{equation} 
One may check by direct computation that $R_\pm e_2(0,n-1) = 
\pm i e_\mp(0,n-1/2\pm 1/2)$ and similarly that $R_\pm 
e_\pm(0,n-1/2\mp 1/2) = -ie_0(0,n)$. Hence, we recover the
structure of the projective cover ${\cal P}_n$.  
\medskip 

The functions on our supergroup carry another (anti-)commuting 
action of the Lie superalgebra $\g$ by left derivations. There 
is a corresponding decomposition which is certainly identical 
to the decomposition in proposition 1. A more interesting 
problem is to decompose the space of functions with respect 
to the graded product $\g \otimes \g$ in which the first factors 
acts through the left regular action while for the second factor
we use the right regular action. The associated decomposition 
is provided by the following proposition. 
\bigskip 

\noindent
{\bf Proposition 2:} (Left-right regular action) {\it 
With respect to the left-right regular action of $\g \otimes 
\g$, the space of functions on the supergroup decomposes 
according to 
$$ L_2(\mbox{\rm GL(1$|$1)}) \ = \ \int_{e\neq 0} de dn \  
   H^L_{\p{-e,-n+1}} \ \otimes\  H^R_{\p{e,n}} \ \oplus  
   \int_0^1 dq \ {\cal J}_q 
   \ \ . 
$$ 
Here ${\cal J}_q, q\in [0,1[,$ denotes a a family of infinite 
dimensional indecomposable representation of $\g \otimes \g$. 
When restricted to either left or right regular action, the 
latter decompose according to 
$$
 \left({\cal J}_{q}\right)_{\g^{R}}  \ \sim \  
   \left({\cal J}_{-q}\right)_{\g^{L}} \ = \ 
    \bigoplus_{a\in \QZ} \ \P_{q+a} 
      \ \ \ \mbox{ for all } \ \ \ q \ \in \ [0,1[\ \ .   
$$}
The first term in the decomposition formula follows from 
proposition 1 as in the case of Lie algebras. Our second 
term involves  unusual infinite dimensional representations
which cannot be further decomposed. They appear as follows. 
We have displayed an 
explicit basis for the 4-dimensional spaces $\P_n$ of the 
right regular action in eq.\ (\ref{Pj}): from this it is 
easy to see that 
$$ L_- : \P_n \ \longrightarrow \ \P_{n+1} \ \ \ \ , \ \ \ \ 
   L_+ : \P_n \ \longrightarrow \ \P_{n-1} \ \ ,  
$$    
i.e.\ that the invariant subspaces $\P_n$ of the right regular 
action are mapped into each other by the left regular action
and vice versa, thus ``linking'' the projectives for left and right 
actions into a big block.  In terms of its decomposition series, the 
structure of $\J_q$ is given by 
\begin{eqnarray*} 
 \J_q: & & \bigoplus_{a \in \QZ} \p{q+a}\otimes \p{-q-a} 
    \\[2mm] & & \hspace*{.5cm} \  \longrightarrow \   
   \bigoplus_{a \in \QZ} \p{q+a+1}\otimes \p{-q-a} 
   \ \oplus \   \bigoplus_{a \in \QZ} \p{q+a}\otimes \p{-q-a-1}  
  \\[2mm]  & & \hspace*{8cm} 
\longrightarrow \   \bigoplus_{a \in \QZ} \p{q+a}\otimes
   \p{-q-a} \ \ . 
\end{eqnarray*} 
Note that $\p{n} \otimes \p{m}$ are atypical 1-dimensional 
representations of $\g \otimes \g$. The rightmost term in 
this filtration of $\J_q$ denotes the so-called socle, i.e.\ 
the largest semi-simple submodule. The leftmost term is the 
head of $\J_q$. It is the largest semi-simple representation 
that arises as a quotient of $J_q$. The term in the middle, 
finally, is the head of the radical.    
\medskip

\subsection{Correlation functions in minisuperspace}

In the minisuperspace theory, fields are represented through 
functions $\phi$ on the supergroup and their correlators are 
computed by integration with the invariant measure, i.e.\ 
$$ \langle \, \prod_{\nu=1}^{m}\ \phi_\nu \, \rangle \ = \ 
  \int d\mu(x,y,\eta_\pm) \ \prod_{\nu=1}^{m}\ 
           \phi_\nu(x,y,\eta_\pm)\ \ . 
$$    
Here we shall be mostly concerned with the correlators 
involving matrix elements of the objects $\vp_\p{e,n}$. 
\smallskip 

In order to prepare for our analysis of correlators of typical
fields we need to introduce a bit of notation. As before, we 
shall denote the eigenstates of $N$ in typical 
representations by $|0\rangle$ and $|1\rangle$. Our choice is 
such that $\Psi^- |0\rangle = 0$. States of an m-fold tensor 
product can be thought of as states in a spin chain of length 
$m$. A basis in this space is given by $|\s_1 \dots \s_m
\rangle$ with $\s_i = 0,1$. In this vector space we shall 
introduce a set of linear maps $E$ by 
$$ E^{\s_1\dots\s_m}_{\s'_1 \dots\s'_m} \ = \ 
   |\s_1\dots\s_m\rangle \ \langle 
    \s'_1\dots\s'_m | \ \ . $$ 
These are the elementary matrices of the state space of 
our spin chain. They will appear later in our formulas 
for the correlation functions of primary fields that 
are associated with typical representations.  
\smallskip 

When we evaluate $m$-point functions of our matrix valued 
functions $\vp_{\p{e,n}}$, we can express the answer in terms 
of the elementary matrices $E$ for a spin chain of length $m$. 
Note that only elementary matrices can arise that preserve 
the number $p$ of spins that are flipped to the position 
$\sigma=1$. We can also observe that an elementary matrix 
with $p$ flipped spins comes multiplied by a delta function 
$\delta(\sum_{\nu=1}^{m} n_\nu -m +p-2)$, i.e.\ an m-point 
function possesses the general form\footnote{The term with 
$p = m$ vanishes due to conservation of the $e$-charge} 
\begin{equation} \label{MSScorr} 
 \langle\, \prod_{\nu=1}^{m} \vp_\p{e_\nu,n_\nu} \, 
   \rangle \ = \ \delta(\sum_{\nu=1}^m e_\nu) \
    \sum_{p=1}^{m-1} \, G_p^{(m)} \, 
    \delta(\sum_{\nu=1}^{m} n_\nu -m +p-2)\ \ , 
\end{equation} 
where $G_p^{(m)}$ are linear combinations of the elementary 
matrices for a spin chain of length $m$ with $p$ spins in 
the $\sigma = 1$ position. We shall find the same structure
for the correlators in the field theory later on. There is 
one more rule, however, that is specific to the particle 
limit: an elementary matrix can only contribute to the 
invariant tensor $G_p^{(m)}$ if it shifts at most one spin 
along the chain. For $m \leq 3$ this conditions is trivially
satisfied, but starting from 4-point functions, some number 
preserving elementary matrices are no longer admitted in the
particle limit. We shall see below that this last condition 
may be violated for the full field theory.

\section{Representation theory of current algebra} 

Based on the experience from the previous section we would now 
like to present a similar analysis of the representation theory 
of the affine algebra. This will ultimately lead us to a 
conjecture on the state space of the GL(1$|$1) WZW model. Our 
proposal will follow closely the outcome of the harmonic 
analysis on the supergroup. The only new ingredient enters 
through an additional spectral flow symmetry of the affine 
algebra.

\subsection{The gl(1$|$1) current algebra} 
Let us begin by listing a few results on the affine algebra 
and its representation theory. The gl(1$|$1) current algebra is 
generated by the modes of two bosonic currents $N(z),E(z)$ and 
two fermionic currents $\Psi^\pm(z)$. Their commutation relations 
read 
\begin{eqnarray}
 [ E_n,N_m] \ = \ k m \delta_{n+m} \ \ \ \ \ &,& \ \ \ \ \  
 [ N_n,\Psi^\pm_m] \ = \ \pm \Psi^\pm_{n+m} \\[2mm]
 \{ \Psi^-_n,\Psi^+_m\} & = & E_{n+m} + km \delta_{n+m} \ \ . 
\end{eqnarray} 
All other (anti-)commutators vanish. This algebra admits an interesting 
family of spectral flow automorphism $\gamma_m$ which acts on generators 
according to 
$$ \gamma_m(E_n) \ = \ E_n + m k \delta_n \ \ \ , \ \ \ 
   \gamma_m(\Psi^\pm_n) \ = \ \Psi^\pm_{n\pm m}
$$ 
and leaves $N_n$ invariant. We shall see later that for integer 
$m$ these automorphisms provide a symmetry of the GL(1$|$1) WZW model. 
\smallskip 

In \cite{Rozansky:1992rx} it was shown that the the Virasoro element 
$L_0$ of the gl(1$|$1) model possesses the following form, 
\begin{eqnarray*} 
 L_0 & = &  \frac{1}{2k}\  (2 N_0 E_0- E_0 + 
       2 \Psi^-_0 \Psi^+_0 +\frac{1}{k} E^2_0) \\[2mm] 
        & & +  \frac{1}{k} \sum_{m\geq0} \, (E_{-m} N_m + 
    N_{-m} E_m + \Psi^-_{-m} \Psi^+_{m} - \Psi^+_{-m} \Psi^-_m
    + \frac{1}{k} E_{-m} E_m)  
\end{eqnarray*} 
Under the action of the spectral flow automorphism, $L_0$ behaves
according to 
$$ \gamma_m (L_0) \ = \ L_0 + m (N_0-1) \ \ . 
$$     
This very simple behavior of $L_0$ plays an important role
in determining the action of the spectral flow on 
representations of the current algebra. 
\smallskip 

\subsection{Representations of the gl(1$|$1) current algebra} 

In the following we shall denote the Verma module over the typical 
representation $\p{e,n}$ by ${\cal V}_\p{e,n}$. As long as $e$ is not 
an integer multiple of the level $k$, the Verma modules are 
irreducible. But when $e = km$, the story is a bit more 
interesting.%
\bigskip%

\noindent 
{\bf Lemma:} {\it The Verma module that is built on the typical 
representations $\p{mk,n}, m \neq 0,$ contains a singular vector 
on the $m^{th}$ level. Explicitly, it is given by 
\begin{equation}    |km,n \pm 1 \rangle_\p{mk,n} \ = \ 
\prod_{p=1}^{|m-1|}\Psi^\mp_{p} \prod_{p=1}^{|m|} \Psi^\pm_{-p} 
|mk,n\rangle \ \ \ \mbox{ for } \ \ \ 0 < \pm m 
       \ \  
\label{sing} 
\end{equation} 
where $|mk,n\rangle$ denotes the ground state of the Verma module 
${\cal V}_\p{mk,n}$.} 
\medskip 

\noindent
{\sc Proof:} Without loss of generality, let us restrict to $m >0$. 
In order to prove our statement we shall begin with the following 
simple fact, 
$$ \Psi^-_q \prod_{p=1}^{m} \Psi^+_{-p} |mk,n\rangle \ = \ 0 
 \ \ \ \mbox{ for all } \ \ q \geq m \ \ . $$
For $q > m$, the formula would hold regardless of the ground state
we use. Only the case $q=m$ is slightly more subtle and it uses that 
$E_0 = mk$. With this insight in mind it is then straightforward 
to establish the Lemma. 
\bigskip 

If we divide the Verma module ${\cal V}_\p{mk,n}$ by the invariant 
subspace that is generated from its singular vector (\ref{sing}) 
we end up with an irreducible representation of the current 
algebra. We shall denote the latter by ${\cal H}_\p{mk,n}$. One 
may also show that the invariant subspace which is built on the 
vector (\ref{sing}) is irreducible and isomorphic to 
${\cal H}_\p{mk,n \pm 1}$. Hence, the  structure of the Verma 
module  ${\cal V}_\p{e,n}$ is encoded in the following diagram 
$$ {\cal V}_\p{mk,n}: \ \ \ \ \ {\cal H}_\p{mk,n} 
   \ \longrightarrow \ {\cal H}_\p{mk,n \pm 1}\ \  \ \ 
   \mbox{ for } \ \ \ 0 < \pm m \ \ . $$ 
For negative $m$, this resembles the structure of the module 
$\p{\overline{0,n}}$. And indeed, one can easily see that the 
Verma  module  ${\cal V}_\p{km,n}$ for $m<0$ is the spectral 
flow image of the Verma module ${\cal V}_{\p{\overline{0,n}}}$.  
Similarly, for positive $m$, the spectral flow takes us from 
the Verma module  ${\cal V}_{\p{0,n+1}}$ to the Verma module
${\cal V}_\p{km,n}$. 
\medskip 

In order to discuss the action of the spectral flow on the 
Verma modules over the projective covers ${\cal P}_n$, we 
need to enlarge the class of affine representations and 
include certain representations $\hat \P_\p{mk,n}$ that are 
known as twisted highest weight modules. These are generated 
from a state $|mk,n\rangle$ satisfying the conditions 
\begin{eqnarray} 
 \Psi^\pm_{r} |mk,n\rangle & = & 0 \ \ \mbox{for} \ \ r > \mp m  
 \label{modedef}
\\[2mm] 
E_r |mk,n\rangle & = & 0 \ = \ N_r|mk,n\rangle  \ \ \mbox{for} 
  \ \ \ r > 0 \ \  \\[2mm] 
E_0 |mk,n\rangle & = & mk |mk,n\rangle \ \ , \ \ 
N_0 |mk,n\rangle \ = \ n |mk,n\rangle
\end{eqnarray} 
by applications of the generators $\Psi^\pm_{\pm r}, r \leq \mp m,$ 
and $E_r,N_r, r > 0$ (note the difference in (\ref{modedef}) 
with the definition of ordinary highest weight modules which would 
involve $r>0$ instead). Let us note that for $m=0$, the construction 
gives us the Verma module of the projective cover $\P_{n}$. We also 
observe that in the twisted highest weight module, the eigenvalues
of $L_0$ are bounded from below, simply because we can only apply a 
finite number of fermionic generators to descend from $|mk,n\rangle$. 
For $m \neq 0$ there are two states of lowest $L_0$ eigenvalue which 
are given by  
$$ \prod_{r=1}^{|m|} \Psi^{- {\rm sign}(m)}_r |mk,n\rangle \ \ \ , 
   \ \ \    
   \prod_{r=0}^{|m|} \Psi^{- {\rm sign}(m)}_r |mk,n\rangle \ \ \ . 
$$ 
For $m\geq 0$ for instance, they possess $L_0$ eigenvalue  $nm -m^2/2 
- m/2$ and transform in the typical representation $\p{mk,n-m}$. 
Consequently, the module 
$\hat \P_\p{mk,n}$ contains ${\cal V}_\p{mk,n-m}$ as a subspace. 
If we divide by the latter we stay with a Verma module ${\cal 
V}_\p{mk,n-m+1}$. This structure in encoded in the following 
diagram, 
\begin{equation}\label{Ppic2}
 \hat \P_\p{mk,n}: \ \ \ \ \ \ \ 
 \xymatrix{ & \H_\p{mk,n-m+2} \ar[dr] &\\
       \H_\p{mk,n-m+1} \ar[dr] \ar[ur] && \H_\p{mk,n-m+1}\ \ . \\
             & \H_\p{mk,n-m} \ar[ur]&}
\end{equation}
It is easy to see that the $\hat \P_\p{mk,n}$ are the images of the 
$\P_{n}$ under the spectral flow.  The picture does not only tell 
us how spectral flow acts on the 
various Verma modules we can built over representations with $E=0$, 
it also displays the twisted highest weight modules $\hat P$ as 
natural cousins of the projective covers $\P_n$. 

\subsection{The state space of the GL(1$|$1) WZW model} 
Having gained some insight into the representation theory of the 
gl(1$|$1) current algebra, it is very tempting to conjecture that 
the state space of the full field theory possesses exactly the 
same structure as the minisuperspace theory with only one extra 
feature: the theory is periodic under shifts of $e$ by 
multiples of the level $k$, 
$$ {\cal H}_{\rm WZW} \ = \ \int_{e \neq m \ {\rm mod} \ k} 
   de dn \  \H^L_{\p{-e,-n+1}} 
   \ \otimes\  \H^R_{\p{e,n}} \ \oplus \  \sum_m \ \int_0^1 dq 
  \ \hat  {\cal J}^{(m)}_q 
   \ \ . 
$$ 
Here $\hat{\cal J}_q, q\in [0,1[,$ denotes a family of indecomposable 
representations of ${\rm \hat gl(1|1)} \otimes {\rm \hat  gl(1|1)}$. 
They are built from irreducible representations according to the 
following diagram 
\begin{eqnarray*} 
 \J^{(m)}_q: & & \bigoplus_{a \in \QZ} \H_\p{mk,q+a}\otimes 
  \H_\p{-mk,-q-a} 
    \\[2mm] & & \hspace*{0cm} \  \longrightarrow \   
   \bigoplus_{a \in \QZ} \H_\p{mk,q+a+1}\otimes \H_\p{-mk,-q-a} 
   \ \oplus \   \bigoplus_{a \in \QZ} \H_\p{mk,q+a}\otimes 
   \H_\p{-mk,-q-a-1}  
  \\[2mm]  & & \hspace*{7.5cm} 
\longrightarrow \   \bigoplus_{a \in \QZ} 
  \H_\p{mk,q+a}\otimes \H_\p{-mk,-q-a} \ \ . 
\end{eqnarray*}     
According to several remarks earlier on, it is clear that 
the Virasoro modes $L_0$ and $\bar L_0$ cannot be diagonalized 
on ${\cal H}_{\rm WZW}$. As we recalled in the introduction, 
such a behavior is closely linked with the existence of 
logarithmic singularities in the operator product expansion
of local fields in the model. 
\smallskip

In the next three sections we shall prove that the state space of the 
WZW model does indeed possess the proposed form. First we shall set up 
a free field approach that will allow us to compute any correlation 
function in the system. Then we use this tool to calculate the 3-point 
couplings. Finally, we shall compute the 4-point functions and show 
that they factorize over the proposed set of states. Along the way 
we shall also prove that the theory possesses the conjectured 
spectral flow symmetry. Needless to add that we will also find
the predicted logarithmic singularities in the correlation 
functions.

\section{Solutions of the full CFT} 
\def\D{{\cal D}} 

Our aim now is to construct correlation function 
of the WZW model on the supergroup GL(1$|$1). We shall obtain 
our explicit formulas through a free field representation 
of the model.

\subsection{Free field theory approach} 

We shall suggest here to think of the WZW model on the supergroup 
GL(1$|$1) as a perturbation of a free field theory. The latter is 
composed from a bosonic linear dilaton background and two chiral 
fermionic bc-systems with central charges $c_\pm = -2$. Our split 
of the theory into this free field theory and an interaction does
not preserve the chiral symmetries of the full model, but it is 
manifestly local. Hence, consistency of our correlation function 
is guaranteed but their relation with the GL(1$|$1) WZW model, and
in particular the chiral symmetry, needs to be established. 
\smallskip  
  
To describe strings that move on the supergroup target GL(1$|$1)
we would normally use the following WZW action 
\begin{equation}
    S_{\rm WZW} \ \ = \ \frac{k}{4\pi} \int_\Sigma d^2z \ 
   \left(\partial X \bar \partial Y + \partial Y \bar \partial X 
    - 2 e^{-iY} \, \bar \partial c_-  \partial c_+ \right) . 
    \label{freeact}
\end{equation}
Here, $X$ and $Y$ are bosonic field on the world-sheet and $c_\pm$ 
are fermionic. The invariant measure on the space of these fields is 
$$ d\mu_{\rm WZW} \ \sim \ \D X \ \D Y \ \D(e^{\frac{-i}{2}Y} c_-) 
         \ \D(e^{\frac{-i}{2}Y} c_+) \ \ . 
$$ 
We shall use another action $S$ to describe this system that 
contains an additional pair of chiral auxiliary fermions $b_\pm$. 
It is built, starting from the following free field theory 
\begin{eqnarray} 
   S_0 & = & \frac{1}{4\pi}\  \int_\Sigma d^2z \ 
   \left( k \, \partial X \bar \partial Y +  
          k \, \partial Y \bar \partial X 
          -  \partial Y \bar \partial Y 
            + Q {\cal R} Y \right) \\[2mm] 
   & & \hspace*{3.5cm} 
       +   \frac{1}{2\pi}\  \int_\Sigma d^2z \ 
                 \left( b_+  \partial c_+ + 
                     b_- \bar \partial c_- \right)\ .  
\end{eqnarray}
Terms in the first line form a linear dilaton background whose 
charge $Q$ we fix to be $Q =  i/2$. Note that the linear 
dilaton term depends only on the $Y$ coordinate. Therefore
the contribution $c^b=2$ of the bosonic fields to the central 
charge in independent of the background charge $Q$. In 
addition, the action $S_0$ contains two chiral bc-systems. We chose 
to have the chiral fields $c_\pm$ with  conformal weight $\Delta^c_\pm  
= 0$ while their partners $b_\pm$ are fields of weight $\Delta^b_\pm = 1$; 
the central charge is then $c_{\pm}^{f}=-2$.
\smallskip 

To the free field theory $S_0$ we now add an interaction term 
of the following simple form 
\begin{equation}
   S(X,Y,b_\pm,c_\pm) \ = \ S_0 + S_{\rm int}  \ = \ S_0 -  
  \frac{1}{\pi k}\  \int_\Sigma d^2z \ e^{iY} \, b_- \, b_+ 
\ \ . 
\end{equation} 
Note that all $X$-independent fields have vanishing conformal 
weight. Hence, the interaction term is massless. In our 
treatment of the theory $S$ we work with the canonical 
measure 
\begin{equation} 
 d\mu \ \sim \ \D X \ \D Y \ \D c_- \ \D c_+\  \D b_- \ \D b_+
 \ \ . 
\end{equation}  
On a formal level it is possible to compare the theory $S$ with 
the original WZW-model on the supergroup GL(1$|$1). The comparison 
makes use of a rather subtle relation between the involved 
measures 
$$ d\mu_{\rm WZW} \ \D(b_-) \ \D( b_+) 
 \ \sim \ \exp\left(\frac{1}{4\pi} \  \int_\Sigma d^2z 
   \left(-  \partial Y \bar \partial Y  + 
    Q {\cal R} Y \right) \right) \ d\mu  \ \ . 
$$ 
Once this is inserted into our theory $S$ , we can integrate 
the auxiliary fields to recover the action of the WZW model. 
\smallskip 

We shall also need a free field representation for the vertex 
operators operators of our theory. Let us begin with the fields
$V_{\p{e,n}}$ that are associated to typical representations. We 
model them after the matrices $\vp_{\p{e,n}}$, i.e.\ 
\begin{equation} \label{Ven} 
 V_{\p{e,n}}(x,\bar x) \ = \ :e^{ieX(x,\bar x)+i(n-1)Y(x,\bar x)}: \ 
   \left(\begin{matrix} 1
    & ic_-(x)  \\ iec_+(\bar x)   & 
    e c_-(x) c_+(\bar x) \end{matrix}\right) \ \ . 
\end{equation} 
Their conformal dimension is given by 
$$ \Delta_{(e,n)} \ = \ (2n-3) \frac{e}{2k} + \frac{e^2}{2k^2}\ \ .
$$ 
Comparison with eq.\ (\ref{vpen}) shows that we have dropped one term 
in the lower right corner of the matrix. When $e\neq 0$, the conformal 
dimension of the omitted vertex operator differs from the dimension of
the other matrix elements so that in some sense (see below) we should 
consider the additional term as `subleading'.  Note that the vertex 
operators (\ref{Ven}) may only be used for $e\neq 0$. We need a new 
prescription to deal with $e=0$. Once more, the expression may be 
motivated with the help of the matrix $\vp_{\p{e,n}}$ which, at $e=0$, 
degenerates to 
$$ \vp_\p{0,n} \ = \ \left(\begin{matrix} e^{i(n-1)y} 
    & i\eta_- e^{i(n-1)y} \\ 0  & 
     e^{iny}  
    \end{matrix}\right)  \ \ \ . 
$$  
Since the lower right corner contains only a single term, there 
is nothing we can omit and consequently the corresponding vertex 
operators are introduced by 
$$ V_\p{0,n}(x,\bar x) \ = \ \left(\begin{matrix} :e^{i(n-1)
    Y(x,\bar x)} & ic_-(x) :e^{i(n-1)Y(x,\bar x)}: \\ 0  & 
     :e^{inY(x,\bar x)}:  
    \end{matrix}\right) \ \ . 
$$
Recall that at $e=0$, there exists a second family of 2-dimensional 
representations $\p{\overline{0,n}}$. The corresponding matrices 
$\overline \vp_\p{0,n}$ of functions on the supergroup are given 
by
$$ \overline\vp_\p{0,n} \ = \ \left(\begin{matrix} e^{iny} 
    & 0  \\ i\eta_+ e^{i(n-1)y} & 
    e^{i(n-1)y}  
    \end{matrix}\right) \ \ .  
$$ 
These may also be obtained from the matrices $\vp_\p{e,n}$, but 
we have to conjugate the latter with the matrix  $W = e \sigma^+ + 
\sigma^-$ ($\sigma^\pm$ are Pauli matrices) before sending $e$ to 
zero. The associated vertex operators are constructed as
$$ \overline V_\p{0,n}(x,\bar x) \ = \ \left(\begin{matrix} 
   :e^{inY(x,\bar x)}: 
    & 0  \\ ic_+(\bar x)  :e^{i(n-1)Y(x,\bar x)}: & 
    :e^{i(n-1)Y(x,\bar x)}:   
    \end{matrix}\right) \ \ .  
$$  
Our matrices $\overline \vp$ do contain the functions $e_+(0,n)$ 
which were not included in $\vp$. Nevertheless, we are still 
missing all functions of the form
$$ \vp_n \ = \ \eta_-\eta_+ e^{i(n-1)y} \ \ . $$ 
The corresponding vertex operators are obtained in the obvious 
way through the formula
$$  V_n(x,\bar x) \ =  \ c_-(x) c_+(\bar x) 
  :e^{i(n-1)Y(x,\bar x)}:\ \ . $$ 
We shall think of the functions $\vp_n$ and the vertex operators 
$V_n$ as being associated with 1-dimensional atypical representations 
of \u11.\footnote{This does not mean that they transform under the 
atypical 1-dimensional representation. In fact, the action of \u11
certainly mixes e.g.\ $\phi_n$ with components of $\phi_\p{0,m}$.} 
Some readers might prefer to construct the additional series of 
objects with the help of projective covers, replacing our $\vp_n$ 
through $4\times 4$ matrices (the representation matrices of the 
supergroup elements). Similarly, vertex operators could then also 
be assembled into $4\times 4$ matrices. At least in the case of the 
\u11 model, such an alternative approach carries no more information 
than the one we have chosen here. Therefore, we shall continue to 
work with single component objects.

\subsection{Computation of correlation functions} 

Let us denote the typical primary fields of the full interacting 
theory $S$ by $\Phi_{\p{e,n}}$. Then our prescription for the 
m-point correlators of this theory is 
\begin{equation} \label{CFTcorr} 
 \langle\, \prod_{\nu=1}^m\,   \Phi_{\p{e_\nu,n_\nu}} 
       (x_\nu,\bar x_\nu) 
\, \rangle \ = \ \sum_{s=0}^\infty\,  \frac{1}{s!}\ 
     \langle \ \left( \frac{-1}{\pi k}\  \int_\Sigma d^2z \ e^{iY} 
      \, b_- \, b_+\right)^s\  \prod_{\nu=1}^m\,    
     V_{\p{e_\nu,n_\nu}} (x_\nu,\bar x_\nu)\  \rangle_0\ \ . 
\end{equation} 
We can employ the same formula if some of the $e_\nu$ vanish 
as long as we agree to insert the corresponding vertex operators 
$V_\p{0,n_\nu}$ for each such primary field into the correlators
on the right hand side. Similarly, the prescription may be used 
to determine correlation functions involving the field theory 
analogue $\Phi_n$ of the functions $\vp_n$ defined above. 
\smallskip  

The correlators on the right hand side are to be evaluated in 
the free field theory $S_0$. In order to determine the latter we
shall use the following simple formula for correlators in the 
bosonic theory
\begin{eqnarray}  
& & \hspace*{-1cm}
   \langle \, \prod_{\nu=1}^{m} V_{(e_\nu,n_\nu)}(x_\nu,\bar x_\nu)\, 
                      \rangle 
\ = \ \prod_{\nu < \mu} |x_\nu-x_\mu|^{-2\alpha_{\nu\mu}} 
\ \delta(\, \sum_{\nu=1}^m \, n_\nu - n-1)\ 
   \delta(\sum_{\nu=1}^m \, e_\nu)   
\\[2mm]
& & \hspace*{1cm} 
\mbox{where} \ \ \ \alpha_{\nu\mu} \ = \ (1-n_\nu)\frac{e_\mu}{k} 
  + (1-n_\mu)\frac{e_\nu}{k} - \frac{e_\nu e_\mu}{k^2} \ \ .    
\end{eqnarray} 
Here, $V_{(e,n)} = :\exp(ieX+i(n-1)Y):$ are standard bosonic 
vertex operators with a somewhat unusual shift in the labels. 
The charge conservation for the e-charge is standard. For the
parameters $n_\nu$, the background charge $Q$ becomes relevant. 
The usual rules tell us that $\sum (n_\nu -1) = 2Q/i = 1$. 
\smallskip 

In addition we will have to evaluate correlation functions in 
the chiral bc-systems. According to the usual rules, non-vanishing 
correlators on the sphere must satisfy $\# c_\pm - \# b_\pm  = 1$, 
i.e.\ the number of insertions of $c_\pm$ must exceed the number 
of insertions of $b_\pm$ by one. For such chiral correlation 
functions one obtains 
\begin{equation} 
 \langle\,  \prod_{\nu=1}^{n} b_-(z_\nu) \ 
           \prod_{\mu=1}^{n+1} c_-(x_\mu) 
  \,  \rangle_0 \ = \  
  \frac{\prod_{\nu < \nu'} (z_\nu-z_{\nu'}) \ \prod_{\mu < \mu'} 
        (x_\mu-x_{\mu'})}{\prod_\nu \prod_\mu (z_\nu -x_\mu)} \ \  
\end{equation}   
and a similar formula applies to $c_+$ and $b_+$. This concludes 
our preparation. We can now turn to a calculation of the 
correlators.   
\smallskip 

It is quite instructive to compare the above expansion 
(\ref{CFTcorr}) for field theory correlators with our  
previous discussion of ``correlation functions'' in 
the minisuperspace limit. From charge conservation 
in the $Y$-direction (parameter $n$) we infer that 
the $s^{th}$ term in the expansion is nonzero if and 
only if 
$$ s + \sum_{\nu=1}^m(n_\nu -1)  = 1 \ \ . $$
Hence, the summation over $s$ in eq.\ (\ref{CFTcorr}) is 
equivalent to the summation over $p$ in our formula 
(\ref{MSScorr}), the precise relation between the two 
summation parameters being $p=s+1$. In section 2.3
we saw that there could only be a finite number of 
terms. The same is true for the full field theory
since the number of $b$-insertions has to be smaller 
that the number of possible $c$-insertions. Hence, 
terms with $s\geq m$ vanish so that the summation 
is finite. A slightly more detailed analysis shows
once more that the term with $s=m-1$ vanishes as 
long as we only insert typical fields. Consequently, 
the last non-vanishing term appears at $p = s+1 = m-1$, 
just as in the minisuperspace model. Let us anticipate, 
however, that correlators of the fields $\Phi_n$
do receive contributions from $s=m-1$.    

\section{The 3-point functions of the GL(1$|$1) model}

For our evaluation of the 3-point functions we shall adopt the 
following strategy. To begin with, we shall construct the 3-point 
functions of the typical fields $\Phi_\p{e,n}$. In the limit where 
$e \rightarrow 0$, these include correlations involving fields 
$\Phi_\p{0,n}$ or $\overline \Phi_\p{0,n}$ so that we do not have 
to list the corresponding 3-point functions separately. All these 
correlators turn out to mimic very closely the minisuperspace 
theory, except from a minor but interesting quantum deformation. 
In a second step, we shall then also determine 3-point functions
involving one or more insertions of the fields $\Phi_n$ that 
come with the functions $\vp_n$. These correlation functions
contain logarithms.  

\subsection{3-point functions of typical fields} 

The first important result of this subsection provides us with 
an explicit formula for the 3-point correlator of typical field. 
It takes the form 
\begin{eqnarray}
\langle \, \Phi_\p{e_1,n_1}(\infty) \, \Phi_\p{e_2,n_2}(1)\,  
           \Phi_\p{e_3,n_3}(x)  \, \rangle 
 & = & \sum_{s=0,1} \ C_s(e_1,e_2,e_3)\ \frac{\langle  
      \vp_\p{e_1,n_1} \, 
     \vp_\p{e_2,n_2} \, \vp_\p{e_3,n_3} \rangle_s}
     {{|1-x|}^{2\Delta_s}} \nonumber   \\[2mm] \nonumber  
& & \hspace*{-7cm} \mbox{ where } \ \ \ C_0(e_1,e_2,e_3) \ = \ 1  
              \ \ \ \mbox{ and } 
      \ \ \ C_1(e_1,e_2,e_3)\  = \ \prod_{i=1}^3                     
      \Gamma(1+\frac{e_i}{k})/\Gamma(1-\frac{e_i}{k})\  
\end{eqnarray} 
The exponents in the denominator are given by $\Delta_0 = (n_2-1) a_3 + 
(n_3-1) a_2 +  a_2 a_3$ and $\Delta_1 = n_2a_3 + n_3a_2 + a_2a_3$ with 
the rescaled parameters $a_i = - e_i/k$. Note that for each choice of 
the parameters $n_i$ at most one of the two terms in the sum can 
contribute. Hence, our result is manifestly consistent with the
conformal invariance of the model. The symbols $\langle f \rangle_s$ 
that appear in the numerator on the right hand side refer to the 
terms in the minisuperspace result for the 3-point function (see
eq.\ (\ref{MSScorr}), i.e.
\begin{eqnarray*} 
 \langle \, \vp_\p{e_1,n_1} \, \vp_\p{e_2,n_2}\,  \vp_\p{e_3,n_3} 
  \, \rangle & = & \!\!\int d\mu\, \vp_\p{e_1,n_1} \, \vp_\p{e_2,n_2}\,  
   \vp_\p{e_3,n_3} \, = \, \sum_{s=0,1} 
  \langle \, \vp_\p{e_1,n_1} \, \vp_\p{e_2,n_2}\,  \vp_\p{e_3,n_3} 
  \, \rangle_s \\[2mm] 
\mbox{where} & & \langle \, \vp_\p{e_1,n_1} \, \vp_\p{e_2,n_2}\,  
   \vp_\p{e_3,n_3}  \, \rangle_s \ = \ 
  G^{(3)}_s \  \delta(\ssum_\nu e_\nu)\, \delta(\ssum_\nu n_\nu-4+s)  
\end{eqnarray*} 
and $G^{(3)}_s$ are the unique invariant tensors in the 
triple tensor products of typical representations. Explicit 
formulas can be worked out by integration of the threefold
products over the supergroup. We shall not need these 
formulas here. Since the terms in our result are proportional
to the tensors $G^{(3)}_s$, our 3-point couplings are manifestly 
gl(1$|$1) covariant. The main difference between the minisuperspace 
limit and the full field theory result arises from the non-trivial, 
e-dependent factor $C_1$ in front of the second term. Note that 
the latter approaches $C_1 \sim 1$ as we send $k$ to infinity, 
thereby reproducing the minisuperspace result in this limit. 
\smallskip 

Our formula for the 3-point functions is not very difficult to 
derive. Note that the first term with $s = 0$ arises from the
corresponding term in the expansion (\ref{CFTcorr}). Since 
there is no screening charge inserted in this case, the 
field theory computation is identical to the associated 
calculation in the minisuperspace theory and hence $C_0=1$. 
As for the second term, the computation is slightly more 
involved. Let us only compute one particular component 
here.    
\begin{eqnarray*} 
 \langle\ ^0_0\Phi_\p{e_1,n_1}(\infty) \, ^1_1\Phi_\p{e_2,n_2}(1)\,  
           ^1_1\Phi_\p{e_3,n_3}(x)  \, \rangle & \sim & 
   \frac{1}{k} \int d^2z e_2 e_3 \ 
     \frac{|z-1|^{2a_2-2}|z-x|^{2a_3-2}}{|1-x|^{2\Delta_0-2}}
   \\[2mm] 
 &  & \hspace{-7.3cm} \ = \ \frac{1}{k} \ \frac{e_2 e_3}{|1-x|^{2\Delta_1}} \ 
       \frac{\Gamma(-a_2) \Gamma(1+a_2+a_3) \Gamma(-a_3)}
        {\Gamma(1+a_3) \Gamma(-a_2-a_3) \Gamma(1+a_2)} 
    \ = \  C_1(e_1,e_2,e_3) \ \frac{e_2+e_3}{|1-x|^{2\Delta_1}}   
\end{eqnarray*}
The $\sim$ in the first line means that we only display the 
coefficient  in front of the $\delta$-functions for the 
contribution with $s=1$ insertion of the interaction, i.e.\ 
we assume implicitly that $e_1 = -e_2+e_3$ and $n_1 = 3-n_2-n_3$.  
In the computation we used a special case of the Dotsenko-Fateev 
integration formula (Appendix A). The other steps are 
straightforward. It is finally easy to see that $e_2+e_3$ 
arises as a result of the corresponding minisuperspace 
computation.  
\smallskip 

It is now rather instructive to study what happens to the 
component $^1_1\Phi_{\p{e,n}}$ as we send $e$ to zero. Since 
we are using the associated vertex operator $ec_-c_+V_{(e,n)}$ 
in the free field computation, one might naively expect that 
the limiting field is zero. Our formula for the 3-point 
coupling, however, shows that this is not the case. Instead 
we find 
$$ \lim_{e\rightarrow 0}\, ^1_1\Phi_{\p{e,n}}\ = \ 
   ^0_0\Phi_{\p{e,n+1}}\ \ . $$
Though this result may appear a bit surprising at first, it 
is actually rather natural. In order to construct the vertex operator 
for typical fields, we had to remove one term from the corresponding 
matrix $\vp_{\p{e,n}}$ of the minisuperspace theory. We declared 
this term to be `subleading' in some sense. But when $e$ is sent 
to zero, the `leading term' in the lower right corner of 
$\vp_{\p{e,n}}$ vanishes so that the other term is no longer 
`negligible'. This is exactly what we may infer from the previous 
formula.      
\smallskip 

Another remark concerns a very interesting new symmetry of the 
field theory that is not present in the minisuperspace theory. 
Note that the coefficients $C_1$ of the 3-point couplings have 
poles whenever one of the $a_i$ becomes a positive integer. 
This behavior seems to distinguish the lines $e = k \mathbb{Z}$ 
in the parameter space of $\p{e,n}$. In the minisuperspace 
theory, only the line $e=0$ was special. Hence, we take the 
behavior of $C_1$ as a first indication that the spectral flow
symmetry might be a symmetry of our physical model, not just 
of its symmetry (see section 3). This is indeed the case. In 
fact, one may show by a short explicit computation that  
$$ \langle \ \Phi_\p{e_1,n_1}(\infty) \, 
   ^0_0\Phi_\p{e_2,n_2}(1)\,  
           ^1_1\Phi_\p{e_3,n_3}(0)  \, \rangle
 \ = \ {\cal N}\, \langle \ \Phi_\p{e_1,n_1}(\infty) \, 
    ^1_1\Phi_\p{e_2+k,n_2-1}(1)\,  
           ^0_0\Phi_\p{e_3-k,n_3+1}(0)  \, \rangle
$$ 
for all the matrix components of $\Phi_\p{e_1,n_1}$. The 
coefficient ${\cal N} = e_3/(e_2+k)$ is due to our 
normalization of the components $^1_1\Phi$ and it would 
be absent had we normalized our fields in the canonical 
way. Consequently, the 3-point functions of typical fields 
are periodic in the parameter $e$ with period length $k$. 
This proves that spectral flow symmetry of the model on 
the level of its 3-point functions functions. 
\smallskip  

Let us finally recall that the 3-point correlators involving 
primaries $\Phi_\p{0,n}$ with $e=0$ may be obtained by taking 
$e_i$ to zero in the above expression of the 3-point couplings. 
In case of $\overline \Phi_\p{0,n}$ one should remember to 
conjugate the typical fields with the matrix $W = e\sigma^+ 
+ \sigma_-$ before performing the limit. All the couplings 
in the resulting correlators agree with the minisuperspace 
result. In fact, when one of the labels $e_i=0$, the other 
two labels differ only by their sign and hence the 
coefficient $C_1 = 1$. 

\subsection{3-point functions involving $\Phi_n$}

It now remains to find correlation functions involving the 
fields $\Phi_n$.\footnote{Recall that the fields $\Phi_n$
and their minisuperspace counterparts $\vp_n$ do not carry 
any matrix indices (see also our comments at the end of 
section 4.1).} These can again be determined by explicit 
computation using our free field representation. Let us start 
by stating the result for a single insertion of the field  
$\Phi_n$, 
\begin{eqnarray} \label{3pt1om}
\langle \,\Phi_{n_1}(\infty) \, \Phi_\p{e_2,n_2}(1)\,  
           \Phi_\p{e_3,n_3}(x)  \, \rangle 
 & = &  \frac{\langle  
      \vp_{n_1} \, 
     \vp_\p{e_2,n_2} \, \vp_\p{e_3,n_3} \rangle}
     {{|1-x|}^{2\Delta}}    \\[2mm] \nonumber  
& &  \hspace*{-2.5cm} + \ \frac{1}{k}\, \frac{\langle  
      \varpi_{n_1} \, 
     \vp_\p{e_2,n_2} \, \vp_\p{e_3,n_3} \rangle}
     {{|1-x|}^{2\Delta}} \, 
  \left({\cal Z} + \vartheta({a_2}) - \log|1-x|^2 \right) \\[2mm] 
\mbox{where} & & \vartheta(a) \ = \ 2\psi(1)-\psi(a) -\psi(1-a)\ \ ,    
\label{vartheta}
\end{eqnarray} 
the function $\varpi_n$ is defined as $\varpi_{n} = \exp (iny)$, 
our symbol $\psi(a)= \Gamma'(a)/\Gamma(a)$ denotes the Di-gamma 
function and the exponent $\Delta$ agrees with the exponents 
$\Delta_0 = \Delta_1$ we introduced previously. Since a 
non-vanishing 3-point coupling requires $a_2 = - a_3$ we obtain 
$\Delta = (n_3-n_2) a_2 - a_2^2$. The constant ${\cal Z}$, finally, 
may be shifted through a field redefinition and therefore its exact 
value is irrelevant. In fact, if we substitute the atypical 
field $\Phi_{n}$ on the left hand side of the above equation by 
$$ \tilde \Phi_n \ = \ \Phi_n + \kappa \cdot \, ^0_0\Phi_\p{0,n+1}$$ 
then the 3-point function remains of the same form with ${\cal Z}$ 
being replaced by  
$$
   \tilde {\cal Z} \ = \ {\cal Z} + \kappa k \ \ . 
$$      
Let us note that with our original definition of the field 
$\Phi_{n}$, the constant ${\cal Z} \sim 1/(e_2+e_3)$ turns
our to be infinite.  
\smallskip 

It is not very difficult to prove the formula (\ref{3pt1om}). 
With the help of our free field representation we find 
\begin{eqnarray*}
\langle \,\Phi_{n_1}(\infty) \, ^1_1\Phi_\p{e_2,n_2}(1)\,  
           ^0_0\Phi_\p{e_3,n_3}(x)  \, \rangle 
 & \sim &  \frac{e_2}{k} \frac{\Gamma(a_2) \Gamma(-\epsilon) 
   \Gamma(1-a_2+\epsilon)} {\Gamma(a_2-\epsilon) \Gamma(1-a_2) 
   \Gamma(1+\epsilon)} \, |1-x|^{2\epsilon}\\[2mm] 
&  & \hspace*{-5cm} =\  \frac{e_2}{k} \left( - \frac{1}{\epsilon} + 
  2 \psi(1) - \psi(a_2) -  \psi(1-a_2) - \log|1-x|^2 + o(\epsilon)\right)      
\end{eqnarray*}
Here we use the same conventions as in the previous subsection
along with some results which may be found in the appendix. 
The expansion around $\epsilon = a_2 + a_3 \sim 0$ is performed 
using  
$$ \Gamma(x+\epsilon)/\Gamma(x) \sim 1 + \epsilon \psi(x) + 
    o(\epsilon^2)\ \ . 
$$ 
In the minisuperspace limit, the associated 3-point function 
assumes the value $e_2$. Hence, we have established eq.\ 
(\ref{3pt1om}) with ${\cal Z} \sim -1/\epsilon$, at least for
one particular component. 
\smallskip   

Similar steps allow us to determine the correlator in case 
there are there two insertions of the field $\Phi_n$, 
 \begin{equation}
\langle \,\Phi_\p{e_1,n_1}(\infty) \, \Phi_{n_2}(1)\,  
           \Phi_{n_3}(x)  \, \rangle 
\  = \  \frac{2}{k} \ \langle   \vp_\p{e_1,n_1} \varpi_{n_2} \, 
    \vp_{n_3} \, \rangle 
 \, \left({\cal Z} + \log|1-x|^2 \right)\ \ . 
\end{equation}   
Here we use the same notation as in eq.\ (\ref{3pt1om}). Note 
that a non-vanishing 3-point coupling with two insertions of 
$\Phi_n$ requires the third field to have $e_1=0$. In the 
minisuperspace correlator on the right hand side we could
also have replaced $\vp_{n_3}$ by $\varpi_{n_3}$ rather 
than performing this substitution on $\vp_{n_2}$.%
\smallskip 
   
Finally, when all three fields are of the type $\Phi_n$, the 3-point 
couplings read  
$$
 \langle \,\Phi_{n_1}(\infty) \, \Phi_{n_2}(1)\,  
           \Phi_{n_3}(x)  \, \rangle 
 \ = \  \frac{1}{k^2}\, \langle   \vp_{n_1} \varpi_{n_2} \, 
       \varpi_{n_3} \, \rangle
 \, \left( 3 {\cal Z}^2 + 2 {\cal Z} \log|1-x|^2 -
 (\log|1-x|^2)^2\right).  
$$
The attentive reader might have noticed that the minisuperspace 
integral on the right hand side contains an infinite factor 
$\int de = \delta (0)$. To understand such a behavior we recall 
that the fields we use are associated with $\delta$-function 
normalizable states. They can all be approached through a series 
of normalizable fields by smearing them with appropriate functions 
in $(e,n)$-space. Correlation functions are finite as long 
as one of the involved fields is normalizable. In our last 
correlator for three fields of the type $\Phi_n$, however, we 
had so set all the parameters $e$ to zero. Hence, from this 
3-point function alone it is no longer possible to deduce any 
finite correlation functions in which at least one field would 
need to be normalizable (and hence to be smeared out in the 
$e$-coordinate). Having uncovered the rather trivial origin 
of the divergence, we shall no longer hesitate to write 
factors $\delta(0)$. 
\smallskip 
 
After this brief digression into mathematical subtleties it is 
instructive to compare our answers for the correlation functions
of symplectic fermions (see e.g.\ \cite{Kausch:2000fu}). The comparison 
shows that our typical vertex operators are very close relatives of the 
twist fields of the symplectic fermion while $\Phi_n$ behave like 
the logarithmic partner of the vacuum in that theory.    

\section{The 4-point function and factorization} 
\def\F{{\cal F}}  
\def\bF{{\bar \F}}
\def\ba{\bar a}
\def\bb{\bar b}
\def\bc{\bar c} 
\def\bs{\bar \s} 
\def\a{\alpha}
\def\e{\varepsilon} 
\def\bz{\bar z} 
\def\bx{\bar x} 
\def\bm{\bar m} 

We are finally in a position to show that the state space we
have proposed at the end of the third section is consistent 
with the factorization of 4-point functions. To this end we
shall now compute at least one special 4-point function of
typical fields and we shall show that it factorizes over 
the conjectures set of possible intermediate states. 
\smallskip 

The 4-point function we are about to compute has the following 
form \cite{Rozansky:1992rx})
$$ G(x,\bar x) \ := \ \langle \, \Phi_{\p{-e'+\e,1-n'}}(\infty) \,
      \Phi_{\p{e,n}}(x,\bar x) \,  \Phi_{\p{-e-\e,1-n}}(1)
       \Phi_{\p{e',n'}}(0)\, 
    \, \rangle \ \ . 
$$  
Let us note that the same 4-point function was also computed in  
\cite{Rozansky:1992rx} as a local solution of the corresponding
Knizhnik-Zamolodchikov equation. Needless to say that our free 
field computations shall give the same answer. In any case, it 
is easy to see that this correlator only receives contributions
from the insertion of one screening charge. In order to spell 
out the result, we rely heavily on the notations that are 
introduced throughout this work (in particular in subsection 
2.3 and in appendix A). Furthermore, it will be convenient to 
work with the rescaled parameters 
$$ a \ = \ - \frac{e}{k} \ \ \ , \ \ \ 
   a' \ = \ - \frac{e'}{k} \ \ \ , \ \ \ 
   \a \ = \ - \frac{\e}{k} \ \ \ . 
$$
With these notations, the four point function $G$ can be written 
in the following form  
\begin{eqnarray} 
G(x,\bar x) & = & \frac{1}{2k}\, 
   |x|^{2\beta} 
 |1-x|^{2\gamma} \ 
{\sum}'_{\s_i,\bs_i} \ E^{\s_1\s_2\s_3\s_4}_{\bs_1\bs_2\bs_3\bs_4}  
 \, {\cal G}(x,\bar x)_{\s_1\s_2\s_3\s_4}^{\bs_1\bs_2\bs_3\bs_4}  
\\[5mm] \nonumber
\mbox{where} \ & & 
  \beta \ =\  (1-n)a' + (1-n')a +aa' 
\\[2mm]\nonumber
  & & \gamma \ =\ (n-1)(\a+a)+na +  a (\a+a)
\end{eqnarray} 
where the summation extends over all spin configurations such that 
$\sum_i \s_i = \sum_i \bs_i = 2$. Note that this includes 
complementary spin assignments, i.e.\ configurations satisfying
$\s_i+\bs_i=1$ for all $i$, which can not arise in the 
minisuperspace theory. It remains to spell out the functions 
${\cal G}$. Up to a sign, they are given by 
\begin{eqnarray*}
{\cal G}(x,\bar x) & = & (-e'+\e)^{\bs_1} e^{\bs_2} (-e-\e)^{\bs_3}
    (e')^{\bs_4} 
\,x^{\s_2\s_4}\,  (x-1)^{\s_2\s_3} \, 
      {\bar x}^{\bs_2\bs_4}\,(\bar x-1)^{\bs_2\bs_3} \ \times 
 \\[2mm] 
 & &  \hspace*{-2cm} \times \ (-1)^{\sum_{\nu\leq \mu \leq 3} \bs_\nu 
   \s_\mu} \ 
 \left[\  |\F(a+\s_2,a'-\a + \s_2+\s_3+\s_4 - 1;
   a+a' + \s_2+\s_4 |x)|^2 \right. \ + \ \\[2mm] 
&  & \hspace*{-2cm} + \left. (-1)^{\s_2+\s_3+\bs_2 + \bs_3} 
   |x^{1-\s_2-\s_4- a-a'}\, 
    \  \F(\s_3-\a-a, 1 - \s_4 - a'; 
  2 -\s_2-\s_4 -a-a'|x)|^2 \right]
\end{eqnarray*}             
Here, the notation $|.|^2$ means that we we multiply the argument 
with an identical factor in which $\F, \s_i$ and $x$ have been 
replaced by the bared quantities. In this form, it is possible 
to compare our result with the expressions that were found in 
\cite{Rozansky:1992rx}. 
\smallskip 

We are now interested in the closed string states that propagate in 
the intermediate channel when $\a \rightarrow 0$. It is convenient 
to rewrite the function ${\cal G}$ at $\a = 0$ in terms of the 
variables $m = 1-\s_2-\s_3$ and $\bm = 1-\bs_2-\bs_3$ which may 
assume values $m,\bm = -1,0,1$, 
\begin{eqnarray*}
{\cal G}(x,\bar x) & = & (-1)^{\bs_1 +\bs_3} (e')^{\bs_1+\bs_4} 
 e^{\bs_2+\bs_3}
 \,  (x-1)^{\s_2\s_3} \,(\bar x-1)^{\bs_2\bs_3} 
 (-1)^{\sum_{\nu\leq \mu \leq 3} \bs_\nu  \s_\mu} \ \times 
 \\[2mm] 
 & &  \hspace{-5mm} \times \ 
 \left[\  |\F(a+\s_2,a' + \s_2+\s_3+\s_4 - 1;
   a+a' + 2\s_2 + \s_3 + \s_4 - 1 + m |x)|^2 \right. \ + \ \\[2mm] 
&  & \hspace*{-1cm} + \left. (-1)^{m +\bm} 
   |x^{1-\s_2-\s_4- a-a'}\, 
    \  \F(\s_3 -a, 1 - \s_4 - a'; 
  1 +\s_3-\s_4 -a-a'+ m |x)|^2 \right]
\end{eqnarray*} 
Since the third argument of the function ${\cal F}$ coincides
with the sum of the first two arguments up to an integer, we
expect logarithms to appear when we expand the correlators
$G$ around $x=1$, 
\begin{eqnarray}\label{4ptatx=1}
G(x,\bar x) & \stackrel{x\rightarrow 1}{\sim}  & 
  \frac{1}{|1-x|^{2\Delta}} \ \left( \phantom{\frac{1}{x}} 
    \hspace*{-3mm} \langle \vp_{\p{-e'+\e,1-n'}} 
   \, \vp_{\p{e,n}} \,  \vp_{\p{-e-\e,1-n}} \vp_{\p{e',n'}}\rangle 
   \right.  \\[2mm] & &  \hspace*{3cm} \left. + \ \frac{ee'}{k}
   \  \delta_{m,0}\,  \delta_{\bm =0}\     
   (\log|1-x|^2 - \vartheta(a) - \vartheta(a'))\right) 
  \ + \ \dots , \nonumber
\end{eqnarray}
where the exponent $\Delta = -\gamma(\a=0)$ is given by $\Delta 
= (1-2n)a-a^2$. The function $\vartheta$ was introduced in eq.\ 
(\ref{vartheta}). 
\medskip 

We would finally like to show that formula (\ref{4ptatx=1}) is 
consistent with the factorization through the set of proposed 
states. The proof employs the following expression for the 
operator product expansion
\begin{eqnarray}
 \Phi_{\p{e,n}}(x) \Phi_{\p{-e,1-n}}(1) & \sim & 
 \frac{1}{|1-x|^{2\Delta}} \left( \int dl \  
\langle \varpi_l \vp_{\p{e,n}} \vp_{\p{-e,1-n}}\rangle \  \Phi_l(1)
  \right. \ + \label{OPE} \\[2mm] 
 & &   \hspace*{-4.7cm}\left. +\,   
   \int dl  \, \langle \vp_l \vp_{\p{e,n}}
  \vp_{\p{-e,1-n}} \rangle \, \Omega_l(1) + \frac{1}{k}\, 
   \langle \varpi_l \vp_{\p{e,n}}
  \vp_{\p{-e,1-n}}\rangle \left(\vartheta(a)-{\cal Z}- 
  \log|1-x|^2 \right)\, \Omega_l(1)
 \right)..  \nonumber
\end{eqnarray} 
where $\Omega_l$ is a shorthand for $\Omega_l = \,^0_0
\Phi_{\p{0,l+1}}$.
This operator product is a direct consequence
of our results on 3-point functions. In order to verify the sign 
in front of the constant ${\cal Z}$ on the right hand side one 
uses that 
$$ \langle \Phi_{n}(\infty) \Phi_{n'}(1)\rangle \ = \ 2 {\cal Z}
\ \delta(0)\, \delta(n+n'-2) \ \ . $$ 
After inserting this operator product into the 4-point correlator, 
we can evaluate the resulting terms with the help of eq.\ (\ref{3pt1om}). 
The first terms in each line of eq.\ (\ref{OPE}) obviously combine into 
the first term of formula (\ref{4ptatx=1}). We may  evaluate the 
contribution from the last term on the right hand side of eq.\ 
(\ref{OPE}) with the help of  
$$    
 \langle \varpi_l \vp_{\p{e,n}}
  \vp_{\p{-e,1-n}}\rangle \ = \ -e \, \delta_{m,0}\, \delta_{\bar m,0} 
   \ \delta(l-1)\ \ . 
$$ 
As before, the formula should be read as a set of equations
for the matrix components $^\sigma_{\bar \sigma}\vp_\p{e,n}$ and 
$^{\sigma'}_{\bar \sigma'}\vp_\p{e,n}$. The quantities $m$ and 
$\bar m$ on the right hand side are defined through $m= 1-\sigma 
- \sigma'$ and $\bar m = 1 - \bar \sigma - \bar \sigma'$. Putting 
all this together we arrive at the second term in formula 
(\ref{4ptatx=1}). Consequently, we have confirmed that our 
4-point functions factorize on the set of states we had  
predicted.

\section{Concluding remarks} 

In this note we have constructed the correlators of the GL(1$|$1) 
WZW model through a free field representation and we have investigated 
some properties of the theory. We have seen in particular that 
some correlators of the model contain logarithmic singularities. 
Let us stress once more that special 4-point functions of this
theory had been computed before \cite{Rozansky:1992rx}. Rozansky
and Saleur had also observed the logarithms which appear 
whenever the intermediate states are associated with 
atypical representations. The new aspect of our approach 
here is that we were able to relate this very closely to 
the geometry (harmonic analysis) of supergroups. To the
best of our knowledge, this is the first time that a family 
of logarithmic conformal field theories comes with a geometric
interpretation. This may well prove to be a valuable source 
for further insights. 
\smallskip 

While our formulas for the 4-point functions agree with those of 
\cite{Rozansky:1992rx}, it is not clear how other aspects relate 
in detail. We notice in particular that in \cite{Rozansky:1992rx}, 
the need to build non trivial knot invariants led to special 
regularizations. These include e.g.\ a prescription to eliminate 
the divergence in the ${\cal Z}$ factors above. In addition, the  
characters of two dimensional representations were required 
to be orthogonal, even though their natural scalar product is 
always zero - as a result, the  metric used in \cite{Rozansky:1992td} 
(formula 148 of that paper) differs by a factor $e$ from the 
invariant metric. In any case, it is very possible that, for 
particular  values of $k$ (especially in the strong quantum 
regime), other consistent quantum theories appear. 
\smallskip 

Another comment concerns an argument in \cite{Gurarieother} which 
suggest that the GL($1|$1) model is a rather trivial example of a 
logarithmic conformal field theory. This assessment is based on 
the observation that its stress energy tensor 
\begin{eqnarray}
    T\ =\ \frac{1}{2k} 
   \left(2NE+\Psi^{-}\Psi^{+}-\Psi^+\Psi^{-}\right)
   +\frac{1}{2 k^{2}}E^{2}
   \end{eqnarray}
is the bottom component of a projective representation under the 
right current algebra, with the top component being 
\begin{eqnarray}
    t\ =\ \frac{1}{8k^{2}} 
      \left(2NE-\Psi^{-}\Psi^{+}+\Psi^+\Psi^{-}\right)
      +\frac{1}{2k} N^{2}\ \  
      \end{eqnarray}
and two intermediate fermionic components of the form 
\begin{eqnarray}
    \{\Psi^{-},t\}\ =\ \frac{1}{2k^{2}}E\Psi^{-}+\frac{1}{2k}
    (N\Psi^{-}+\Psi^{-}N)\nonumber\\[2mm]
    \{\Psi^{+},t\}\ =\ -\frac{1}{2k^{2}}E\Psi^{+}-\frac{1}{2k}
        (N\Psi^{+}+\Psi^{+}N)
        \end{eqnarray}
The  operator product expansion between the bosonic 
components $t$ and $T$,  
\begin{eqnarray}
    T(z)t(w)&=&\frac{1}{2k}{\frac{1}{(z-w)^{4}}}+\frac{2t(w)}{ 
    (z-w)^{2}}+\frac{\partial t}{z-w}\label{tTOPE} 
\end{eqnarray}
can be used to argue very easily that $L_0$ is diagonalizable on 
this multiplet, a conclusion which is not in contradiction with 
anything we have said before since the identification between 
$L_0$ and the Casimir element applies only to highest weight states. 
Looking at this one multiplet alone, it would seem that more 
``interesting''  logarithmic theories \cite{Gurarieother} are 
those for which the Virasoro field $T$ appears together with its
partner $t$ on the right hand side of the  operator product 
\ (\ref{tTOPE}). In fact, the action of $L_0$ in the 
Virasoro multiplet of such theories ceases to be diagonalizable 
Let us stress, however, that the GL(1$|$1) WZW model is much 
richer than this observation would suggest. As our results 
show, it possesses many  multiplets with non-diagonalizable 
$L_0$ e.g.\ even within the space of ground states. The top 
component of another potentially interesting multiplet can be  
obtained with $L_{-2}$ on $\Phi_{1}$. 
\smallskip 

Even though our analysis here was carried out for GL(1$|$1),  we do 
not expect the results to be much different for compactified U(1$|$1)
model. In the latter case, the spectrum of $e$ and $n$ should be 
discrete, and, in the full quantum field theory, winding will have 
to be introduced.

Irrespective of whether we choose U(1$|$1) or GL(1$|$1), we note that 
the spectrum of the theory is not bounded from below. This is expected 
since the gl(1$|$1) metric is not positive definite - a fact manifest, 
for instance, in that the naive functional integral for the free field 
representation say (\ref{freeact}) is divergent. This feature is generic 
of supergroups, and it was suggested by Zirnbauer in particular 
\cite{Zirnbauer:1999ua} that the WZW model could only be defined by 
trading the target space for a Riemannian symmetric superspace with 
real submanifold $H^{1}\times S^{1}$. We have not followed that route 
here, observing instead that quantum mechanics on GL(1$|$1) was well 
defined, and assuming that there existed a quantum field theory 
reducing to it in the minisuperspace limit. 
\smallskip

Let us finally point out that the geometric arguments that lead to 
the existence of indecomposables in the spectrum were not specific 
to the particular model under consideration. All they required 
was the presence of a Lie-superalgebra symmetry and the 
existence of the identity field in the spectrum of the theory. 
The latter always sits in an atypical representation and is - 
at least whenever the theory contains a typical field multiplet 
- part of a larger indecomposable projective representation.
The existence of an identity field also has a rather simple 
geometric origin: it appears for all theories in which the 
bosonic manifold of the target space is compact. For 
non-compact target spaces, the identity can only be part 
of the spectrum if it may be approximated by normalizable 
functions. This is the case for flat target spaces, i.e.\ 
in the example we have studied. In more generic non-compact
curved backgrounds, however, the identity is separated by 
a gap from the normalizable states of the theory. We 
therefore conclude that models with a compact (or flat) 
target space and a Lie superalgebra symmetry provide  
examples of logarithmic conformal field theory. This 
is certainly a vast class.  
\smallskip

The insights of this note might  be relevant  also for 
non-compact backgrounds once we admit world-sheets with 
boundaries. In geometric terms, the boundary conditions
we impose along the various  boundary components are 
interpreted as branes. Such branes wrap certain subsets
of the target space which may be either non-compact 
or compact. In the latter case, the boundary spectrum 
does contain an identity field even if the bulk 
spectrum does not. For branes that preserve some
Lie superalgebra symmetries we are therefore back 
with a setup that resembles the one we discussed 
in the previous paragraph. Therefore we expect to 
find logarithmic singularities in the boundary 
correlators of a compact brane theory. We plan  
to come back to such issues in a forthcoming 
publication.  
\bigskip

\noindent  
{\bf Acknowledgment:} We wish to thank Anne Taormina for 
her participation in the early stages of this project and
Thomas Quella for numerous comments for pointing out 
several mistakes in earlier versions of this text. We 
also acknowledge useful discussions/exchange with Michael 
Flohr, Matthias Gaberdiel, Gerhard G\"otz, Andreas Ludwig, Nick Read, Sylvain 
Ribault and Didina Serban. VS would like to thank the
organizers of the string theory programmes at the Fields 
Institute, Toronto, and the KITP in Santa Barbara for 
their hospitality during the final stage of the work. 
This research was partially supported by the EU Research 
Training Network grants ``Euclid'', contract number 
HPRN-CT-2002-00325, ``ForcesUniverse'', contract 
number MRTN-CT-2004-005104, by the National Science 
Foundation under Grant No. PHY99-0794, and by the Department of Energy.

\section{Appendix A: Some integral formulas}
\def\bz{{\bar z}} 
\setcounter{equation}{0}

Correlation functions can be computed from the free field 
representation using the following simple consequence of the 
Dotsenko-Fateev integral formula,   
\begin{eqnarray*} 
 && \hspace*{-1cm} \frac{1}{\pi} \int d^2z \ z^{a} \bz^{\ba} 
   (1-z)^{b}(1-\bz)^{\bb} (z-x)^{c} (\bz-\bx)^{\bc} 
      \\[2mm] 
 &=&\!|\F(-c,-c-1-a-b;-c-a|x)|^2   
    + (-1)^{c-\bc+b-\bb} | x^{a+c+1} \F (-b,a+1;a+c+2|x)|^2
\ . \end{eqnarray*} 
We expressed the result of the integration through the
following functions  
\begin{eqnarray} 
\F(a,b;c|x) & = &  \frac{\Gamma(c-b)\Gamma(b)}{\Gamma(c)} 
  \ _2F_1(a,b;c|x) \ \ ,\label{F} \\[2mm] 
\bF(\ba,\bb;\bc|\bar x) & = & 
\frac{\Gamma(1-\bc)}{\Gamma(1-\bc+\bb)\Gamma(1-\bb)} \ 
  _2F_1(\ba,\bb;\bc|\bar x) \ \ . \label{bF}      
\end{eqnarray} 
Validity of the integration  formula requires that all the 
differences $a-\ba, b-\bb$ and $c-\bc$ are integers. When 
one pair of exponents, e.g.\ the labels $a,\bar a$, 
vanishes, then the result simplifies to 
\begin{equation} 
 \hspace*{-1cm} \frac{1}{\pi} \int d^2z \  
   (1-z)^{b}(1-\bz)^{\bb} (z-x)^{c} (\bz-\bx)^{\bc} 
  \ = \ |\F(-c,-c-1-b;-c|x)|^2 \ \ . 
\end{equation} 
This integral formula is used frequently in our evaluation 
of the 3-point couplings. For generic values of $b,c$ we 
have 
\begin{equation}  
 |\F(-c,-c-1-b;-c|x)|^2 \ = \ 
 \frac{\Gamma(1+b) \Gamma(-1-c-b) \Gamma(1+\bar c)} 
  {\Gamma(-c) \Gamma(-\bar b) \Gamma(2+\bar c + \bar b)}
\ |1-x|^{2+b+c+\bar b + \bar c}\ \ . 
\end{equation}

\end{document}